\documentclass[manuscript,sigconf,dvipsnames]{acmart}

\acmVolume{0}
\acmNumber{0}
\acmArticle{0}
\acmMonth{0}

\copyrightyear{2022} 
\acmYear{2022} 
\setcopyright{rightsretained} 
\acmConference[RecSys '22]{Sixteenth ACM Conference on Recommender Systems}{September 18--23, 2022}{Seattle, WA, USA}
\acmBooktitle{Sixteenth ACM Conference on Recommender Systems (RecSys '22), September 18--23, 2022, Seattle, WA, USA}
\acmDOI{10.1145/3523227.3547405}
\acmISBN{978-1-4503-9278-5/22/09}

\usepackage{amsmath}
\usepackage{soul}
\setuldepth{Berlin}

\graphicspath{{figure/}{figures/}}

\colorlet{LineColorA}{RoyalBlue}
\colorlet{ShadeColorA}{RoyalBlue!20!White}

\colorlet{LineColorB}{Orange}
\colorlet{ShadeColorB}{Orange!20!White}

\colorlet{LineColorC}{Gray}
\colorlet{ShadeColorC}{Gray!20!White}

\colorlet{LineColorC}{PineGreen}
\colorlet{ShadeColorC}{PineGreen!20!White}

\colorlet{LineColorD}{Dandelion}
\colorlet{ShadeColorD}{Dandelion!20!White}

\colorlet{LineColorE}{Violet}
\colorlet{ShadeColorE}{Violet!20!White}

\colorlet{LineColorF}{Salmon}
\colorlet{ShadeColorF}{Salmon!20!White}

\begin{document}

\title[Merlin HugeCTR]{%
Merlin HugeCTR: GPU-accelerated Recommender System Training and Inference%
}


\author{Joey Wang}
\email{zehuanw@nvidia.com}
\orcid{0000-0002-1072-2651}
\affiliation{%
  \institution{NVIDIA}
  \city{Beijing}
  \country{China}
}
\author{Yingcan Wei}
\email{yingcanw@nvidia.com}
\orcid{0000-0002-5093-7382}
\affiliation{%
  \institution{NVIDIA}
  \city{Shanghai}
  \country{China}
}
\author{Minseok Lee}
\email{minseokl@nvidia.com}
\orcid{0000-0002-8367-1939}
\affiliation{%
  \institution{NVIDIA}
  \city{Seoul}
  \country{South Korea}
}
\author{Matthias Langer}
\email{mlanger@nvidia.com}
\orcid{0000-0003-1776-8000}
\affiliation{%
  \institution{NVIDIA}
  \city{Shanghai}
  \country{China}
}
\author{Fan Yu}
\email{fayu@nvidia.com}
\orcid{0000-0001-8454-3923}
\affiliation{%
  \institution{NVIDIA}
  \city{Shanghai}
  \country{China}
}
\author{Jie Liu}
\email{kingsleyl@nvidia.com}
\orcid{0000-0002-4293-4827}
\affiliation{%
  \institution{NVIDIA}
  \city{Shanghai}
  \country{China}
}
\author{Alex Liu}
\email{aleliu@nvidia.com}
\orcid{0000-0001-6111-7040}
\affiliation{%
  \institution{NVIDIA}
  \city{Beijing}
  \country{China}
}
\author{Daniel Abel}
\email{dabel@nvidia.com}
\orcid{0000-0003-1592-5981}
\affiliation{%
  \institution{NVIDIA}
  \city{Shanghai}
  \country{China}
}
\author{Gems Guo}
\email{gemsg@nvidia.com}
\orcid{0000-0003-3786-9888}
\affiliation{%
  \institution{NVIDIA}
  \city{Beijing}
  \country{China}
}
\author{Jianbing Dong}
\email{jianbingd@nvidia.com}
\orcid{0000-0002-1910-6417}
\affiliation{%
  \institution{NVIDIA}
  \city{Beijing}
  \country{China}
}
\author{Jerry Shi}
\email{jershi@nvidia.com}
\orcid{0000-0003-1446-0326}
\affiliation{%
  \institution{NVIDIA}
  \city{Shanghai}
  \country{China}
}
\author{Kunlun Li}
\email{kunlunl@nvidia.com}
\orcid{0000-0001-7762-8478}
\affiliation{%
  \institution{NVIDIA}
  \city{Shanghai}
  \country{China}
}

\renewcommand{\shortauthors}{J. Wang \emph{et al.}}

\begin{abstract}
    In this talk, we introduce Merlin HugeCTR. Merlin HugeCTR is an open source, GPU-accelerated integration framework for click-through rate estimation. It optimizes both training and inference, whilst enabling model training at scale with model-parallel embeddings and data-parallel neural networks. In particular, Merlin HugeCTR combines a high-performance GPU embedding cache with an hierarchical storage architecture, to realize low-latency retrieval of embeddings for online model inference tasks. In the MLPerf v1.0 DLRM model training benchmark, Merlin HugeCTR achieves a speedup of up to 24.6x on a single DGX A100 (8x A100) over PyTorch on 4x4-socket CPU nodes (4x4x28 cores). Merlin HugeCTR can also take advantage of multi-node environments to accelerate training even further. Since late 2021, Merlin HugeCTR additionally features a hierarchical parameter server (HPS) and supports deployment via the NVIDIA Triton server framework, to leverage the computational capabilities of GPUs for high-speed recommendation model inference. Using this HPS, Merlin HugeCTR users can achieve a 5\textasciitilde62x speedup (batch size dependent) for popular recommendation models over CPU baseline implementations, and dramatically reduce their end-to-end inference latency.
\end{abstract}



\maketitle
\setlength{\fboxsep}{0pt}%
\newcounter{tmp}%

\section{High Performance Training}\label{s:intro}
\begin{figure*}[tb]
    \begin{minipage}[t]{0.475\hsize}%
        \centering%
        \includegraphics[width=\hsize]{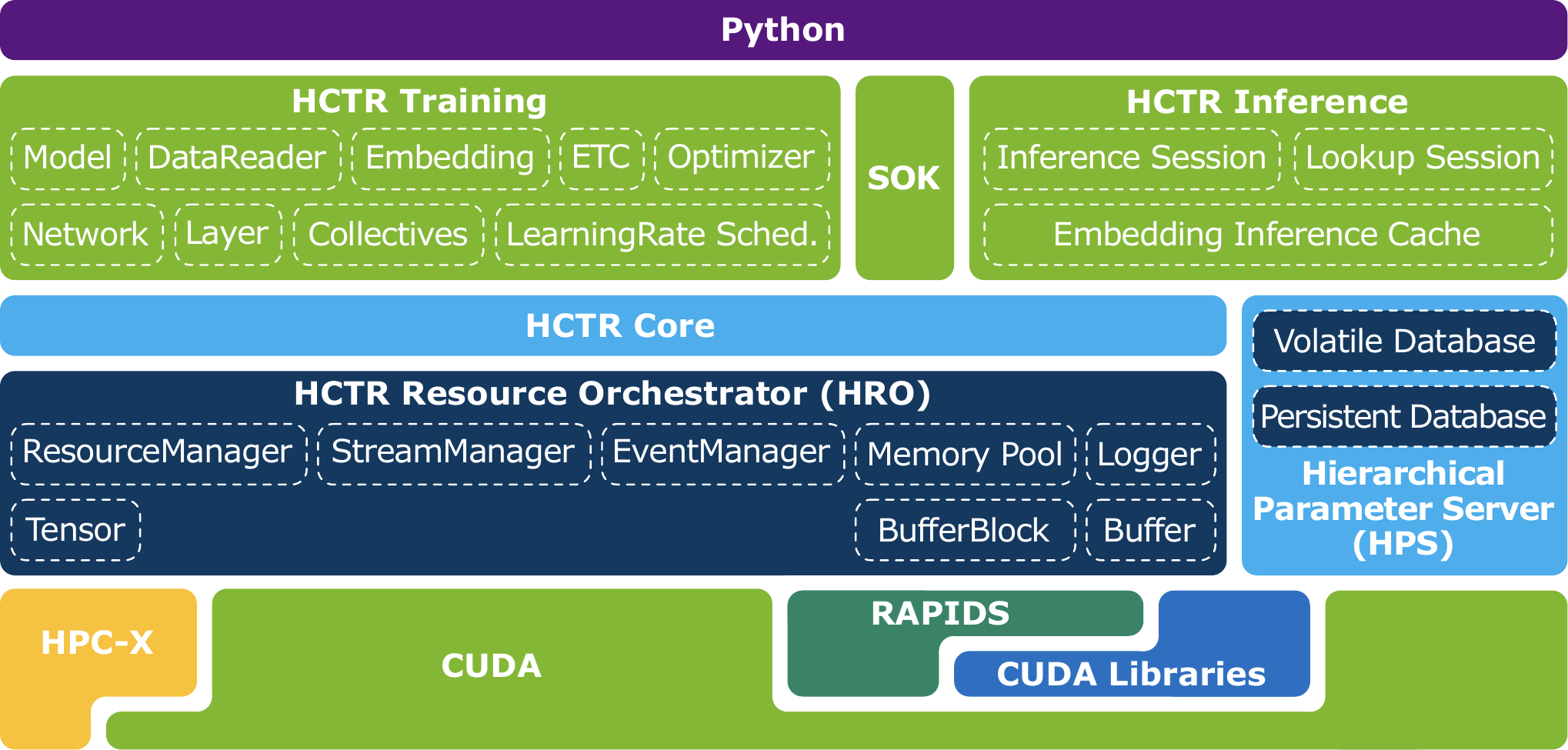}%
        \caption{HugeCTR Architecture}%
        \label{f:hugectr-arch}%
    \end{minipage}%
    \hfill%
    \begin{minipage}[t]{0.475\hsize}%
        \centering%
        \includegraphics[width=\hsize]{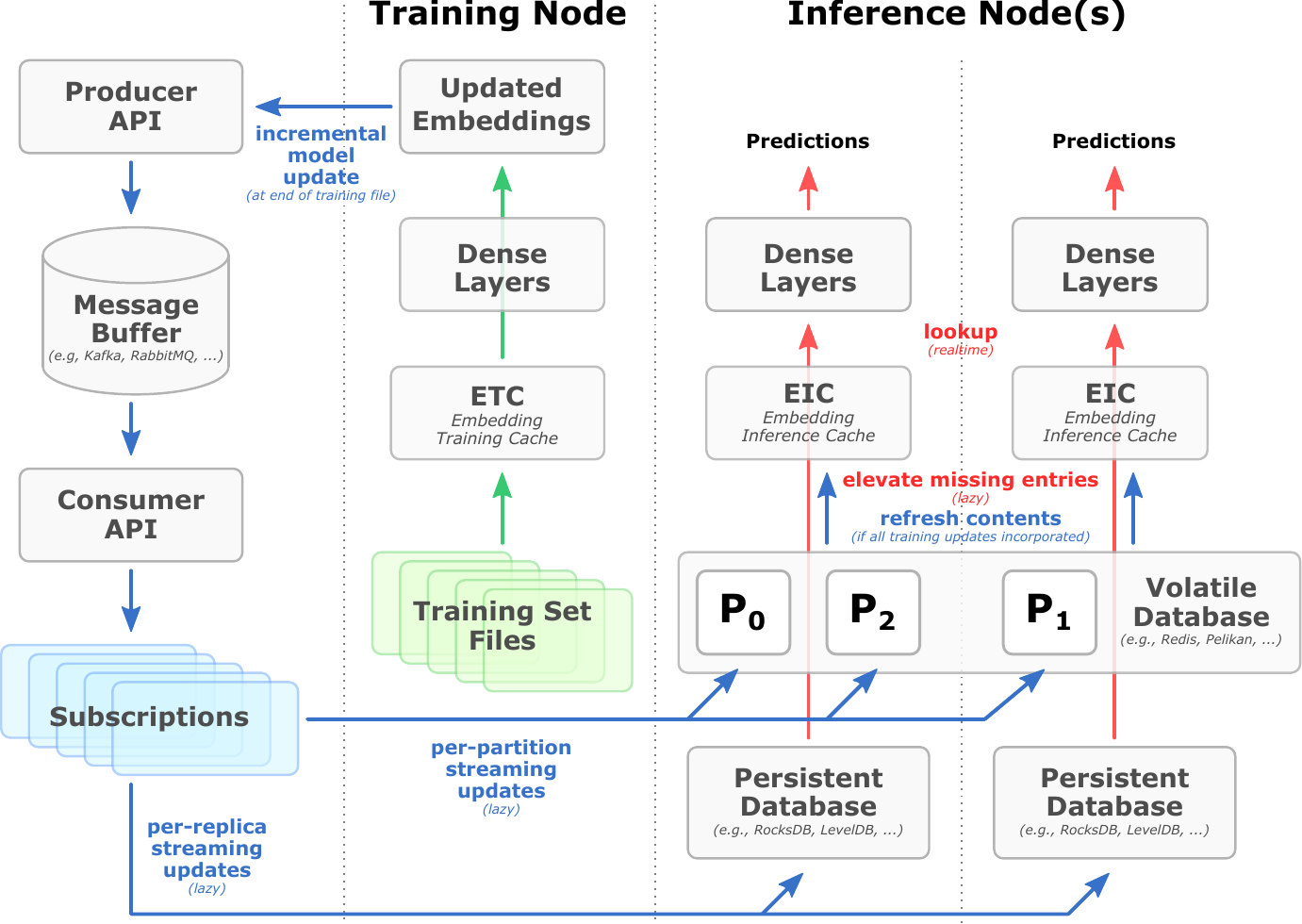}%
        \caption{General data-flow in HugeCTR model deployments.}%
        \label{f:general-data-flow}%
    \end{minipage}%
\end{figure*}

Being a component of the open source package NVIDIA Merlin, HugeCTR is designed for accelerating recommendation system workloads, such as click-through rate (CTR) estimation, on NVIDIA GPUs (cf. \autoref{f:hugectr-arch}). It provides a GPU-accelerated parallel processing optimized framework to distribute model training across multiple GPUs and cluster nodes. In particular, HugeCTR implements model parallelism (MP) for embedding tables, and data parallelism (DP) \cite{DeepRecSys,ScaleFreeCTR,GeePS,Clipper,Angel,Distributed_Hierarchical,FlexPS,Parameter_Hub,Fast_Distributed_Training} for popular recommendation models and their variants including Wide and Deep Learning (WDL; \cite{Wide_Deep}), Deep Cross Network (DCN; \cite{DCN}), DeepFM \cite{DeepFM}, and Deep Learning Recommendation Model (DLRM; \cite{dlrm}). 

\textbf{GPU-based MP and DP training}. 
HugeCTR provides native support for both MP and DP training \cite{bennun2019dist-deep-learning,langer2020ddls} to facilitate the needs of large embedding tables. To balance the load best and achieve maximum performance, the categorical features and resulting embedding tables need to be placed across GPUs. The ideal placement can vary depending on the characteristics of the input data and the model hyper-parameters. To embrace a variety of use cases, HugeCTR supports three different embedding layer types:
\begin{itemize}
    \item \ul{\emph{Localized slot embedding hash}}: All embeddings that belong to the same slot (or table) are stored in the same GPU. This type of embedding layer is suitable if each categorical feature belongs to a slot that can fit into the memory of a single GPU. When doing intra-slot reductions for multi-hot features, no inter-GPU communication is required. After each local reduction step, all-to-all communication is conducted to share the results of the lookup operation across all available GPUs along the batch dimension.
    
    \item \ul{\emph{Distributed slot embedding hash}}: Each GPU only keeps a shard of the embedding table (MP). Therefore, the total size of an embedding feature can exceed the memory capacity of a single GPU. Embeddings are distributed to the respective GPUs according to the hash value of the feature. Consequently, the available storage capacity and communication demand between the GPUs increases proportionally with the number of participating GPUs.
    
    \item \ul{\emph{Hybrid sparse embedding}} is a key technique for achieving industry leading performance in large-scale recommendation model training with NVIDIA GPUs \cite{kanter2021mlperf-1-1-results}. It combines DP and MP for maximum performance. When doing forward or backward propagation, a local cache is used to speedup access for high frequency embeddings and avoid the need for communication between GPUs (DP). For low frequency embeddings, HugeCTR allows utilizing memory across all available GPUs to realize load-balanced sharded embedding feature storage (MP). The all-to-all communication pattern is used to exchange embedding vectors between GPUs.
\end{itemize}
\textbf{Online training}. 
Our Embedding Training Cache (ETC) allows training large models up to terabyte size, by loading subsets of embedding tables on-demand into the GPU. 
\emph{E.g.}, assume you have a cluster consisting of 2 nodes. Each node is equipped with four A100 80~GB GPUs. Using only the GPUs, you could theoretically train models that are at most 640~GB (8x 80~GB) large. With the ETC, you can train models that exceed this constraint. Moreover, our ETC realizes efficient incremental training by carefully balancing accuracy and performance. Notable features of the ETC are:
\begin{enumerate}
    \item \ul{\emph{Broad support}}: Suitable for most single-node multi-GPU and multi-node multi-GPU configurations.
    \item \ul{\emph{Staged-PS}}: Allows scaling embedding tables up to the combined host memory sizes of all nodes.
    \item \ul{\emph{Cached-PS}}: Allows scaling embedding tables up to the capacity of the hard-disk or a Network File System (NFS).
    \item \ul{\emph{Continuous / online training}}: Updated embedding features can be retrieved during training. To enable online training, we support deploying these updates to inference parameter servers in real-time.
\end{enumerate}
\textbf{Multi-node training} makes it easy to train embedding tables of arbitrary size. Multi-node HugeCTR deployments achieve near DP execution performance by distributing the sparse portion of the model (\emph{i.e.}, the embedding layer) across all nodes, while replicating dense model parts (\emph{e.g.}, a DNN) in each GPU. Scalable high-speed inter- and intra-node communication is realized through leveraging NCCL (\emph{i.e.}, the NVIDIA Collective Communications Library; \cite{NCCL}).

\textbf{Mixed precision training} increases computational throughput, while simultaneously reducing the model's memory footprint. For example, to save memory bandwidth and capacity, we can store 
model parameters 
in compressed or truncated data formats like FP16. When processing such compressed parameters, HugeCTR takes advantage of TensorCores to boost the performance of downstream model layers 
while retaining most of the numerical precision.

\section{Easy to use Python interfaces}\label{s:easy}
HugeCTR is a major building brick of NVIDIA Merlin \cite{Merlin}, which is an application framework and ecosystem that facilitates all phases\textemdash{}from experimentation to production\textemdash{}of recommender system development with GPU acceleration. To extend HugeCTR’s functionality and accessibility, we provide a series of features that are easy to use and integrate with other platforms (\emph{e.g.}, Python-based high-level interfaces, such as the TensorFlow Python wrapper, \emph{etc}.).

\textbf{Sparse Operation Kit (SOK)} is a Python package that exposes HugeCTR's GPU-accelerated operations for sparse model training. The package is designed to be compatible with common deep learning frameworks such as TensorFlow \cite{abadi2015tensorflow}. SOK is compatible with the DP training strategies provided by common synchronized training frameworks, such as Horovod \cite{sergeev2017horovod} and TensorFlow's \emph{distribute strategy}. Models built with SOK embedding layers automatically take advantage of DP. Additional \emph{DP-to-MP} and \emph{MP-to-DP} transformations are needed when SOK is used to scale up DNN models from a single to multiple GPUs.

\textbf{Keras-like Python API}. HugeCTR also offers high level abstracted Python APIs for both recommender system training and inference. To lower the entry bar for new users, the design of these APIs follows the \emph{look \& feel} of the popular deep learning framework Keras. Thereby, the tedious task of deploying individual training and inference jobs in an optimized manner on a specific hardware topology can be delegated to HugeCTR, so that practitioners can focus on their overall AI algorithm design.
 
\textbf{HugeCTR to ONNX Converter} is a Python package for converting HugeCTR models to the open-source AI model storage format ONNX (=Open Neural Network Exchange; \cite{linux-foundation2019onnx}). Thus, HugeCTR models are implicitly compatible with and can be loaded using other ONNX-compatible deep learning frameworks, such as PyTorch \cite{paszke2019pytorch} or TensorFlow.

\section{Hierarchical Parameter Server for Inference}\label{s:hps}
The Hierarchical Parameter Server (HPS) is HugeCTR's mechanism for extending the space available for embedding storage beyond the constraints of GPUs using various memory resources from across the inference cluster. It enables doing inference for models with huge embedding tables. HPS is implemented as a 3-level hierarchical cache architecture that utilizes GPU GDDR and/or high-bandwidth memory (HBM), distributed CPU memory and local SSD storage resources. The communication mechanisms between these components ensure that the most frequently used embeddings reside in the GPU embedding cache, while somewhat regularly occurring embeddings are cached in CPU memory. A full copy of all model parameters, including those that rarely occur, is retained on the hard-disk/SSD of each node. To minimize delays, we overlap parameter updating and the migration of missing parameters from higher storage levels (SSD \textcolor{Red}{$\rightarrow$} CPU memory \textcolor{Red}{$\rightarrow$} GPU memory; \emph{cf.} red data-flow graph in \autoref{f:general-data-flow}) with dense model computations.

The \textbf{embedding inference cache} (level 1) is a dynamic cache designed for recommendation model inference. It attempts to improve the lookup performance of recommendation models by reducing additional/repetitive parameter movement through cleverly utilizing data locality to keep the frequently used features (\emph{i.e.} the hot features) in the GPU memory. Our cache features several optimized query and manipulation operators, as well as a dynamic insertion, and an asynchronous refresh mechanism to retain a high cache hit rate during online inference.

\textbf{Volatile database (VDB}; level 2\textbf{)} layers utilize volatile memory resources, that require traversal through a NVLink or the PCIe bus to access them from a GPU (\emph{e.g.}, system memory), to store partial copies of the embedding parameters. They act as an extension to the GPU embedding cache and are queried if an embedding is recalled that is currently not present in the GPU memory. In comparison to GPU memory, system memory can be extended at lower costs. To grow even further, VDBs can be spread across the system memories of multiple nodes in an inference cluster. For example, our Redis VDB template implementation allows using distributed Redis instances \cite{redis_cluster} as a storage backend for embeddings.

\textbf{Persistent database (PDB}; level 3\textbf{)} layers use hard-disks/SSDs to permanently store entire embedding tables (\emph{i.e.}, all model parameters). To that end, PDBs are regarded as a slow but virtually inexhaustible extra storage area. They are helpful in improving the prediction accuracy for datasets that exhibit an extreme long-tail distribution, with a large number of embeddings that contribute valuable information to an embedding model, but appear too rarely, so that caching them in GPU/CPU memory is ineffective. Each PDB instance can serve as backup and ultimate ground truth for any number of models. Key collisions are avoided by forming separate key namespaces for each embedding table.

\textbf{Online model updating}. If an incremental update has been applied to some embedding table entries, either during online training (=frequent/incremental updates), or after completing an offline training (\emph{cf.} \autoref{s:intro}), the latest versions of the respectively updated embeddings have to be propagated to all inference nodes. Our HPS achieves this functionality using a dedicated online updating mechanism. The blue [\textcolor{blue}{$\rightarrow$}] data-flow graph in \autoref{f:general-data-flow} illustrates this process. First, the training nodes dump their updates to an Apache Kafka-based message buffer \cite{sax2018kafka}. This is done via our \emph{Message Producer API}, which handles serialization, batching, and the organization of updates into distinct message queues for each embedding table. Inference nodes that have loaded the affected model can use the corresponding \emph{Message Source API} to discover and subscribe to these message queues. Received updates are then subsequently applied to the respective local VDB shards and the PDB. The GPU embedding cache polls its associated VDB/PDB for updates and replaces embeddings if necessary. This refresh cycle is configurable to best fit the training schedule. When using online training, the GPU embedding cache periodically (\emph{e.g.}, every $n$ minutes, hours, \emph{etc}.) scans for updates and refresh its contents. During offline training, poll-cycles are instigated by the Triton model management API \cite{triton_model_management}. 

\section{Future Work}
For the next version of HugeCTR, we currently develop a novel next-generation embedding technique that is cross-framework, user friendly and offers high performance. It will support fusing embedding tables with different embedding vector size/combiner, and permit both dense and sparse inputs. Thus, it will provide users with the flexibility to add arbitrary lookup types into one embedding. Our next-generation embedding will feature a cross-framework backend, 
that can autonomously determine what embedding table placements grant optimal performance.

We also continue to extend HPS with additional features, including but not limited to better support for next generation GPU innovations, customized plugins for more target platforms, efficient embedding compression/quantization techniques, and further optimizing performance with flexibility and usability considered.

\section{Authors}
All the authors are developers from NVIDIA’s Developer Technology Group in the APAC region. They are mainly working on analyzing and optimizing the end-to-end performance of GPU applications. Based on their expertise, they are actively developing HugeCTR to make it more accessible to a wider range of developers.

\bibliographystyle{ACM-Reference-Format}
\bibliography{references}


\begin{thebibliography}{25}


\ifx \showCODEN    \undefined \def \showCODEN     #1{\unskip}     \fi
\ifx \showDOI      \undefined \def \showDOI       #1{#1}\fi
\ifx \showISBNx    \undefined \def \showISBNx     #1{\unskip}     \fi
\ifx \showISBNxiii \undefined \def \showISBNxiii  #1{\unskip}     \fi
\ifx \showISSN     \undefined \def \showISSN      #1{\unskip}     \fi
\ifx \showLCCN     \undefined \def \showLCCN      #1{\unskip}     \fi
\ifx \shownote     \undefined \def \shownote      #1{#1}          \fi
\ifx \showarticletitle \undefined \def \showarticletitle #1{#1}   \fi
\ifx \showURL      \undefined \def \showURL       {\relax}        \fi
\providecommand\bibfield[2]{#2}
\providecommand\bibinfo[2]{#2}
\providecommand\natexlab[1]{#1}
\providecommand\showeprint[2][]{arXiv:#2}

\bibitem[Abadi et~al\mbox{.}(2016)]%
        {abadi2015tensorflow}
\bibfield{author}{\bibinfo{person}{Mart{\'\i}n Abadi}, \bibinfo{person}{Paul
  Barham}, \bibinfo{person}{Jianmin Chen}, \bibinfo{person}{Zhifeng Chen},
  \bibinfo{person}{Andy Davis}, \bibinfo{person}{Jeffrey Dean},
  \bibinfo{person}{Matthieu Devin}, \bibinfo{person}{Sanjay Ghemawat},
  \bibinfo{person}{Geoffrey Irving}, \bibinfo{person}{Michael Isard},
  \bibinfo{person}{Manjunath Kudlur}, \bibinfo{person}{Josh Levenberg},
  \bibinfo{person}{Rajat Monga}, \bibinfo{person}{Sherry Moore},
  \bibinfo{person}{Derek~G. Murray}, \bibinfo{person}{Benoit Steiner},
  \bibinfo{person}{Paul Tucker}, \bibinfo{person}{Vijay Vasudevan},
  \bibinfo{person}{Pete Warden}, \bibinfo{person}{Martin Wicke},
  \bibinfo{person}{Yuan Yu}, {and} \bibinfo{person}{Xiaoqiang Zheng}.}
  \bibinfo{year}{2016}\natexlab{}.
\newblock \showarticletitle{{TensorFlow: A System for Large-Scale Machine
  Learning}}. In \bibinfo{booktitle}{\emph{12th USENIX Symposium on Operating
  Systems Design and Implementation (OSDI 16)}}. \bibinfo{publisher}{USENIX
  Association}, \bibinfo{address}{Savannah, GA, USA},
  \bibinfo{pages}{265--283}.
\newblock
\showISBNx{978-1-931971-33-1}
\urldef\tempurl%
\url{https://www.usenix.org/conference/osdi16/technical-sessions/presentation/abadi}
\showURL{%
\tempurl}


\bibitem[Ben-Nun and Hoefler(2019)]%
        {bennun2019dist-deep-learning}
\bibfield{author}{\bibinfo{person}{Tal Ben-Nun} {and} \bibinfo{person}{Torsten
  Hoefler}.} \bibinfo{year}{2019}\natexlab{}.
\newblock \showarticletitle{{Demystifying Parallel and Distributed Deep
  Learning: An In-Depth Concurrency Analysis}}.
\newblock \bibinfo{journal}{\emph{ACM Comput. Surv.}} \bibinfo{volume}{52},
  \bibinfo{number}{4}, Article \bibinfo{articleno}{65} (\bibinfo{date}{aug}
  \bibinfo{year}{2019}), \bibinfo{numpages}{43}~pages.
\newblock
\showISSN{0360-0300}
\urldef\tempurl%
\url{https://doi.org/10.1145/3320060}
\showDOI{\tempurl}


\bibitem[Cheng et~al\mbox{.}(2016)]%
        {Wide_Deep}
\bibfield{author}{\bibinfo{person}{Heng-Tze Cheng}, \bibinfo{person}{Levent
  Koc}, \bibinfo{person}{Jeremiah Harmsen}, \bibinfo{person}{Tal Shaked},
  \bibinfo{person}{Tushar Chandra}, \bibinfo{person}{Hrishi Aradhye},
  \bibinfo{person}{Glen Anderson}, \bibinfo{person}{Greg Corrado},
  \bibinfo{person}{Wei Chai}, \bibinfo{person}{Mustafa Ispir},
  \bibinfo{person}{Rohan Anil}, \bibinfo{person}{Zakaria Haque},
  \bibinfo{person}{Lichan Hong}, \bibinfo{person}{Vihan Jain},
  \bibinfo{person}{Xiaobing Liu}, {and} \bibinfo{person}{Hemal Shah}.}
  \bibinfo{year}{2016}\natexlab{}.
\newblock \showarticletitle{{Wide \& Deep Learning for Recommender Systems}}.
  In \bibinfo{booktitle}{\emph{Proceedings of the 1st Workshop on Deep Learning
  for Recommender Systems}} (Boston, MA, USA) \emph{(\bibinfo{series}{DLRS
  2016})}. \bibinfo{publisher}{Association for Computing Machinery},
  \bibinfo{address}{New York, NY, USA}, \bibinfo{pages}{7--10}.
\newblock
\showISBNx{9781450347952}
\urldef\tempurl%
\url{https://doi.org/10.1145/2988450.2988454}
\showDOI{\tempurl}


\bibitem[Crankshaw et~al\mbox{.}(2017)]%
        {Clipper}
\bibfield{author}{\bibinfo{person}{Daniel Crankshaw}, \bibinfo{person}{Xin
  Wang}, \bibinfo{person}{Guilio Zhou}, \bibinfo{person}{Michael~J. Franklin},
  \bibinfo{person}{Joseph~E. Gonzalez}, {and} \bibinfo{person}{Ion Stoica}.}
  \bibinfo{year}{2017}\natexlab{}.
\newblock \showarticletitle{{Clipper: A Low-Latency Online Prediction Serving
  System}}. In \bibinfo{booktitle}{\emph{14th USENIX Symposium on Networked
  Systems Design and Implementation (NSDI 17)}}. \bibinfo{publisher}{USENIX
  Association}, \bibinfo{address}{Boston, MA}, \bibinfo{pages}{613--627}.
\newblock
\showISBNx{978-1-931971-37-9}
\urldef\tempurl%
\url{https://www.usenix.org/conference/nsdi17/technical-sessions/presentation/crankshaw}
\showURL{%
\tempurl}


\bibitem[Cui et~al\mbox{.}(2016)]%
        {GeePS}
\bibfield{author}{\bibinfo{person}{Henggang Cui}, \bibinfo{person}{Hao Zhang},
  \bibinfo{person}{Gregory~R. Ganger}, \bibinfo{person}{Phillip~B. Gibbons},
  {and} \bibinfo{person}{Eric~P. Xing}.} \bibinfo{year}{2016}\natexlab{}.
\newblock \showarticletitle{{GeePS: Scalable Deep Learning on Distributed GPUs
  with a GPU-Specialized Parameter Server}}. In
  \bibinfo{booktitle}{\emph{Proceedings of the 11th European Conference on
  Computer Systems}} (London, United Kingdom) \emph{(\bibinfo{series}{EuroSys
  '16})}. \bibinfo{publisher}{Association for Computing Machinery},
  \bibinfo{address}{New York, NY, USA}, Article \bibinfo{articleno}{4},
  \bibinfo{numpages}{16}~pages.
\newblock
\showISBNx{9781450342407}
\urldef\tempurl%
\url{https://doi.org/10.1145/2901318.2901323}
\showDOI{\tempurl}


\bibitem[Goodwin et~al\mbox{.}(2021)]%
        {triton_model_management}
\bibfield{author}{\bibinfo{person}{David Goodwin} {et~al\mbox{.}}}
  \bibinfo{year}{2021}\natexlab{}.
\newblock \bibinfo{title}{Triton Inference Server}.
\newblock
  \bibinfo{howpublished}{\url{https://github.com/triton-inference-server/server/blob/main/docs/model\_management.md\#model-control-mode-explicit}}.
\newblock
\newblock
\shownote{Accessed: 2022-05-15}.


\bibitem[Guo et~al\mbox{.}(2021)]%
        {ScaleFreeCTR}
\bibfield{author}{\bibinfo{person}{Huifeng Guo}, \bibinfo{person}{Wei Guo},
  \bibinfo{person}{Yong Gao}, \bibinfo{person}{Ruiming Tang},
  \bibinfo{person}{Xiuqiang He}, {and} \bibinfo{person}{Wenzhi Liu}.}
  \bibinfo{year}{2021}\natexlab{}.
\newblock \bibinfo{booktitle}{\emph{{ScaleFreeCTR: MixCache-Based Distributed
  Training System for CTR Models with Huge Embedding Table}}}.
\newblock \bibinfo{publisher}{Association for Computing Machinery},
  \bibinfo{address}{New York, NY, USA}, \bibinfo{pages}{1269--1278}.
\newblock
\showISBNx{9781450380379}
\urldef\tempurl%
\url{https://doi.org/10.1145/3404835.3462976}
\showDOI{\tempurl}


\bibitem[Guo et~al\mbox{.}(2017)]%
        {DeepFM}
\bibfield{author}{\bibinfo{person}{Huifeng Guo}, \bibinfo{person}{Ruiming
  Tang}, \bibinfo{person}{Yunming Ye}, \bibinfo{person}{Zhenguo Li}, {and}
  \bibinfo{person}{Xiuqiang He}.} \bibinfo{year}{2017}\natexlab{}.
\newblock \showarticletitle{{DeepFM: A Factorization-Machine Based Neural
  Network for CTR Prediction}}. In \bibinfo{booktitle}{\emph{Proceedings of the
  26th International Joint Conference on Artificial Intelligence}} (Melbourne,
  Australia) \emph{(\bibinfo{series}{IJCAI'17})}. \bibinfo{publisher}{AAAI
  Press}, \bibinfo{address}{Palo Alto, CA, USA}, \bibinfo{pages}{1725--1731}.
\newblock
\showISBNx{9780999241103}
\urldef\tempurl%
\url{https://doi.org/10.5555/3172077.3172127}
\showDOI{\tempurl}


\bibitem[Gupta et~al\mbox{.}(2020)]%
        {DeepRecSys}
\bibfield{author}{\bibinfo{person}{Udit Gupta}, \bibinfo{person}{Samuel Hsia},
  \bibinfo{person}{Vikram Saraph}, \bibinfo{person}{Xiaodong Wang},
  \bibinfo{person}{Brandon Reagen}, \bibinfo{person}{Gu-Yeon Wei},
  \bibinfo{person}{Hsien-Hsin~S. Lee}, \bibinfo{person}{David Brooks}, {and}
  \bibinfo{person}{Carole-Jean Wu}.} \bibinfo{year}{2020}\natexlab{}.
\newblock \showarticletitle{{DeepRecSys: A System for Optimizing End-To-End
  At-Scale Neural Recommendation Inference}}. In
  \bibinfo{booktitle}{\emph{ACM/IEEE 47th Annual International Symposium on
  Computer Architecture (ISCA)}}. \bibinfo{publisher}{IEEE Press},
  \bibinfo{address}{Valencia, Spain}, \bibinfo{pages}{982--995}.
\newblock
\urldef\tempurl%
\url{https://doi.org/10.1109/ISCA45697.2020.00084}
\showDOI{\tempurl}


\bibitem[Gupta et~al\mbox{.}(2021)]%
        {Fast_Distributed_Training}
\bibfield{author}{\bibinfo{person}{Vipul Gupta}, \bibinfo{person}{Dhruv
  Choudhary}, \bibinfo{person}{Peter Tang}, \bibinfo{person}{Xiaohan Wei},
  \bibinfo{person}{Xing Wang}, \bibinfo{person}{Yuzhen Huang},
  \bibinfo{person}{Arun Kejariwal}, \bibinfo{person}{Kannan Ramchandran}, {and}
  \bibinfo{person}{Michael~W. Mahoney}.} \bibinfo{year}{2021}\natexlab{}.
\newblock \showarticletitle{{Training Recommender Systems at Scale:
  Communication-Efficient Model and Data Parallelism}}. In
  \bibinfo{booktitle}{\emph{Proceedings of the 27th ACM SIGKDD Conference on
  Knowledge Discovery \& Data Mining}} (Virtual Event, Singapore)
  \emph{(\bibinfo{series}{KDD '21})}. \bibinfo{publisher}{Association for
  Computing Machinery}, \bibinfo{address}{New York, NY, USA},
  \bibinfo{pages}{2928--2936}.
\newblock
\showISBNx{9781450383325}
\urldef\tempurl%
\url{https://doi.org/10.1145/3447548.3467080}
\showDOI{\tempurl}


\bibitem[Huang et~al\mbox{.}(2018)]%
        {FlexPS}
\bibfield{author}{\bibinfo{person}{Yuzhen Huang}, \bibinfo{person}{Tatiana
  Jin}, \bibinfo{person}{Yidi Wu}, \bibinfo{person}{Zhenkun Cai},
  \bibinfo{person}{Xiao Yan}, \bibinfo{person}{Fan Yang},
  \bibinfo{person}{Jinfeng Li}, \bibinfo{person}{Yuying Guo}, {and}
  \bibinfo{person}{James Cheng}.} \bibinfo{year}{2018}\natexlab{}.
\newblock \showarticletitle{{FlexPS: Flexible Parallelism Control in Parameter
  Server Architecture}}.
\newblock \bibinfo{journal}{\emph{Proc. VLDB Endow.}} \bibinfo{volume}{11},
  \bibinfo{number}{5} (\bibinfo{year}{2018}), \bibinfo{pages}{566--579}.
\newblock
\showISSN{2150-8097}
\urldef\tempurl%
\url{https://doi.org/10.1145/3177732.3177734}
\showDOI{\tempurl}


\bibitem[Jiang et~al\mbox{.}(2018)]%
        {Angel}
\bibfield{author}{\bibinfo{person}{Jie Jiang}, \bibinfo{person}{Lele Yu},
  \bibinfo{person}{Jiawei Jiang}, \bibinfo{person}{Yuhong Liu}, {and}
  \bibinfo{person}{Bin Cui}.} \bibinfo{year}{2018}\natexlab{}.
\newblock \showarticletitle{{Angel: A new large-scale machine learning
  system}}.
\newblock \bibinfo{journal}{\emph{National Science Review}}
  \bibinfo{volume}{5} (\bibinfo{year}{2018}), \bibinfo{pages}{216--236}.
\newblock
\showISSN{2095-5138}
\urldef\tempurl%
\url{https://doi.org/10.1093/nsr/nwx018}
\showDOI{\tempurl}


\bibitem[Kanter et~al\mbox{.}(2021)]%
        {kanter2021mlperf-1-1-results}
\bibfield{author}{\bibinfo{person}{David Kanter}, \bibinfo{person}{Peter
  Mattson}, {et~al\mbox{.}}} \bibinfo{year}{2021}\natexlab{}.
\newblock \bibinfo{title}{{ML$\cdot$Commons / MLperf v1.1 Results}}.
\newblock
  \bibinfo{howpublished}{\url{https://mlcommons.org/en/training-normal-11}}.
\newblock
\newblock
\shownote{Accessed: 2022-03-15}.


\bibitem[Langer et~al\mbox{.}(2020)]%
        {langer2020ddls}
\bibfield{author}{\bibinfo{person}{Matthias Langer}, \bibinfo{person}{Zhen He},
  \bibinfo{person}{Yanbo Xue}, {and} \bibinfo{person}{Wenny Rahayu}.}
  \bibinfo{year}{2020}\natexlab{}.
\newblock \showarticletitle{{Distributed Training of Deep Learning Models: A
  Taxonomic Perspective}}.
\newblock \bibinfo{journal}{\emph{IEEE Transactions on Parallel and Distributed
  Systems}} \bibinfo{volume}{31}, \bibinfo{number}{12} (\bibinfo{year}{2020}),
  \bibinfo{pages}{2802--2818}.
\newblock
\showISSN{2331-8422}
\urldef\tempurl%
\url{https://doi.org/10.1109/TPDS.2020.3003307}
\showDOI{\tempurl}


\bibitem[Luo et~al\mbox{.}(2018)]%
        {Parameter_Hub}
\bibfield{author}{\bibinfo{person}{Liang Luo}, \bibinfo{person}{Jacob Nelson},
  \bibinfo{person}{Luis Ceze}, \bibinfo{person}{Amar Phanishayee}, {and}
  \bibinfo{person}{Arvind Krishnamurthy}.} \bibinfo{year}{2018}\natexlab{}.
\newblock \showarticletitle{{Parameter Hub: A Rack-Scale Parameter Server for
  Distributed Deep Neural Network Training}}. In
  \bibinfo{booktitle}{\emph{Proceedings of the ACM Symposium on Cloud
  Computing}} (Carlsbad, CA, USA) \emph{(\bibinfo{series}{SoCC '18})}.
  \bibinfo{publisher}{Association for Computing Machinery},
  \bibinfo{address}{New York, NY, USA}, \bibinfo{pages}{41--54}.
\newblock
\showISBNx{9781450360111}
\urldef\tempurl%
\url{https://doi.org/10.1145/3267809.3267840}
\showDOI{\tempurl}


\bibitem[Naumov et~al\mbox{.}(2019)]%
        {dlrm}
\bibfield{author}{\bibinfo{person}{Maxim Naumov}, \bibinfo{person}{Dheevatsa
  Mudigere}, \bibinfo{person}{Hao{-}Jun~Michael Shi}, \bibinfo{person}{Jianyu
  Huang}, \bibinfo{person}{Narayanan Sundaraman}, \bibinfo{person}{Jongsoo
  Park}, \bibinfo{person}{Xiaodong Wang}, \bibinfo{person}{Udit Gupta},
  \bibinfo{person}{Carole{-}Jean Wu}, \bibinfo{person}{Alisson~G. Azzolini},
  \bibinfo{person}{Dmytro Dzhulgakov}, \bibinfo{person}{Andrey Mallevich},
  \bibinfo{person}{Ilia Cherniavskii}, \bibinfo{person}{Yinghai Lu},
  \bibinfo{person}{Raghuraman Krishnamoorthi}, \bibinfo{person}{Ansha Yu},
  \bibinfo{person}{Volodymyr Kondratenko}, \bibinfo{person}{Stephanie Pereira},
  \bibinfo{person}{Xianjie Chen}, \bibinfo{person}{Wenlin Chen},
  \bibinfo{person}{Vijay Rao}, \bibinfo{person}{Bill Jia},
  \bibinfo{person}{Liang Xiong}, {and} \bibinfo{person}{Misha Smelyanskiy}.}
  \bibinfo{year}{2019}\natexlab{}.
\newblock \showarticletitle{{Deep Learning Recommendation Model for
  Personalization and Recommendation Systems}}.
\newblock \bibinfo{journal}{\emph{CoRR}}  \bibinfo{volume}{abs/1906.00091}
  (\bibinfo{year}{2019}), \bibinfo{numpages}{10}~pages.
\newblock
\urldef\tempurl%
\url{http://arxiv.org/abs/1906.00091}
\showURL{%
\tempurl}


\bibitem[NVIDIA(2016)]%
        {NCCL}
\bibfield{author}{\bibinfo{person}{NVIDIA}.} \bibinfo{year}{2016}\natexlab{}.
\newblock \bibinfo{title}{{The NVIDIA Collective Communications Library
  (NCCL)}}.
\newblock \bibinfo{howpublished}{\url{https://developer.nvidia.com/nccl}}.
\newblock
\newblock
\shownote{Accessed: 2022-04-15}.


\bibitem[NVIDIA(2020)]%
        {Merlin}
\bibfield{author}{\bibinfo{person}{NVIDIA}.} \bibinfo{year}{2020}\natexlab{}.
\newblock \bibinfo{title}{{Merlin: A Framework for Building High-Performance,
  Deep learning-based Recommender Systems.}}
\newblock \bibinfo{howpublished}{\url{ https://developer.nvidia.com/nvidia-
  merlin}}.
\newblock
\newblock
\shownote{Accessed: 2022-05-15}.


\bibitem[open~source project(2021)]%
        {redis_cluster}
\bibfield{author}{\bibinfo{person}{Redis open~source project}.}
  \bibinfo{year}{2021}\natexlab{}.
\newblock \bibinfo{title}{{Redis Cluster}}.
\newblock
  \bibinfo{howpublished}{\url{https://redis.io/docs/manual/scaling\#creating-and-using-a-redis-cluster}}.
\newblock
\newblock
\shownote{Accessed: 2022-05-15}.


\bibitem[Paszke et~al\mbox{.}(2019)]%
        {paszke2019pytorch}
\bibfield{author}{\bibinfo{person}{Adam Paszke}, \bibinfo{person}{Sam Gross},
  \bibinfo{person}{Francisco Massa}, \bibinfo{person}{Adam Lerer},
  \bibinfo{person}{James Bradbury}, \bibinfo{person}{Gregory Chanan},
  \bibinfo{person}{Trevor Killeen}, \bibinfo{person}{Zeming Lin},
  \bibinfo{person}{Natalia Gimelshein}, \bibinfo{person}{Luca Antiga},
  \bibinfo{person}{Alban Desmaison}, \bibinfo{person}{Andreas K\"{o}pf},
  \bibinfo{person}{Edward Yang}, \bibinfo{person}{Zach DeVito},
  \bibinfo{person}{Martin Raison}, \bibinfo{person}{Alykhan Tejani},
  \bibinfo{person}{Sasank Chilamkurthy}, \bibinfo{person}{Benoit Steiner},
  \bibinfo{person}{Lu Fang}, \bibinfo{person}{Junjie Bai}, {and}
  \bibinfo{person}{Soumith Chintala}.} \bibinfo{year}{2019}\natexlab{}.
\newblock \showarticletitle{{PyTorch: An Imperative Style, High-Performance
  Deep Learning Library}}. In \bibinfo{booktitle}{\emph{Proceedings of the 33rd
  International Conference on Neural Information Processing Systems}}.
  \bibinfo{publisher}{Curran Associates Inc.}, \bibinfo{address}{Red Hook, NY,
  USA}, Article \bibinfo{articleno}{721}, \bibinfo{numpages}{12}~pages.
\newblock
\urldef\tempurl%
\url{https://doi.org/10.5555/3454287.3455008}
\showDOI{\tempurl}


\bibitem[Sax(2018)]%
        {sax2018kafka}
\bibfield{author}{\bibinfo{person}{Matthias~J. Sax}.}
  \bibinfo{year}{2018}\natexlab{}.
\newblock \bibinfo{booktitle}{\emph{{Apache Kafka}}}.
\newblock \bibinfo{publisher}{Springer International Publishing},
  \bibinfo{address}{Cham}, \bibinfo{pages}{1--8}.
\newblock
\showISBNx{978-3-319-63962-8}
\urldef\tempurl%
\url{https://doi.org/10.1007/978-3-319-63962-8\_196-1}
\showDOI{\tempurl}


\bibitem[Sergeev and Del~Balso(2017)]%
        {sergeev2017horovod}
\bibfield{author}{\bibinfo{person}{Alex Sergeev} {and} \bibinfo{person}{Mike
  Del~Balso}.} \bibinfo{year}{2017}\natexlab{}.
\newblock \bibinfo{title}{Meet Horovod: Uber’s Open Source Distributed Deep
  Learning Framework for TensorFlow}.
\newblock \bibinfo{howpublished}{\url{https://eng.uber.com/horovod}}.
\newblock
\newblock
\shownote{Accessed: 2022-05-15}.


\bibitem[{The Linux Foundation}(2019)]%
        {linux-foundation2019onnx}
\bibfield{author}{\bibinfo{person}{{The Linux Foundation}}.}
  \bibinfo{year}{2019}\natexlab{}.
\newblock \bibinfo{title}{{Open Neural Network Exchange}}.
\newblock \bibinfo{howpublished}{\url{https://onnx.ai}}.
\newblock
\newblock
\shownote{Accessed: 2022-05-15}.


\bibitem[Wang et~al\mbox{.}(2017)]%
        {DCN}
\bibfield{author}{\bibinfo{person}{Ruoxi Wang}, \bibinfo{person}{Bin Fu},
  \bibinfo{person}{Gang Fu}, {and} \bibinfo{person}{Mingliang Wang}.}
  \bibinfo{year}{2017}\natexlab{}.
\newblock \showarticletitle{{Deep \& Cross Network for Ad Click Predictions}}.
  In \bibinfo{booktitle}{\emph{Proceedings of the ADKDD'17}} (Halifax, NS,
  Canada) \emph{(\bibinfo{series}{ADKDD'17})}. \bibinfo{publisher}{Association
  for Computing Machinery}, \bibinfo{address}{New York, NY, USA}, Article
  \bibinfo{articleno}{12}, \bibinfo{numpages}{7}~pages.
\newblock
\showISBNx{9781450351942}
\urldef\tempurl%
\url{https://doi.org/10.1145/3124749.3124754}
\showDOI{\tempurl}


\bibitem[Zhao et~al\mbox{.}(2020)]%
        {Distributed_Hierarchical}
\bibfield{author}{\bibinfo{person}{Weijie Zhao}, \bibinfo{person}{Deping Xie},
  \bibinfo{person}{Ronglai Jia}, \bibinfo{person}{Yulei Qian},
  \bibinfo{person}{Ruiquan Ding}, \bibinfo{person}{Mingming Sun}, {and}
  \bibinfo{person}{Ping Li}.} \bibinfo{year}{2020}\natexlab{}.
\newblock \showarticletitle{{Distributed Hierarchical GPU Parameter Server for
  Massive Scale Deep Learning Ads Systems}}.
\newblock \bibinfo{journal}{\emph{Proceedings of Machine Learning and Systems}}
   \bibinfo{volume}{2} (\bibinfo{year}{2020}), \bibinfo{pages}{412--428}.
\newblock
\urldef\tempurl%
\url{https://proceedings.mlsys.org/paper/2020/file/f7e6c85504ce6e82442c770f7c8606f0-Paper.pdf}
\showURL{%
\tempurl}


\end{thebibliography}


\end{document}